\documentclass[a4paper,12pt]{article}
\pdfoutput=1
\usepackage{epsfig,amsmath,amsfonts,amsthm}
\usepackage[figuresright]{rotating}
\usepackage[left=28mm,right=20mm,top=30mm,bottom=20mm]{geometry}
\usepackage{subfig}
\usepackage{graphicx}
\usepackage{rotating}
\usepackage{booktabs}
\usepackage{moreverb}

\DeclareGraphicsExtensions{.pdf,.png,.jpg}
\usepackage{natbib}
\numberwithin{equation}{section}

\def\e{\hbox{E}}

\def\var{\hbox{Var}}

\def\cv{\hbox{CV}}

\def\e{\hbox{E}}

\def\Bin{\hbox{Bin}}

\def\var{\hbox{Var}}

\def\cv{\hbox{CV}}

\def\logit{\hbox{logit}}

\usepackage[nokeyprefix]{refstyle}
\usepackage{varioref}
\usepackage{xr}
\usepackage{hyperref}

\begin{document}
\title{
Exploring  Consequences of Simulation Design for  Apparent Performance of Statistical Methods.\\
2: Results from simulations with normally and uniformly distributed sample sizes
}

\author{Elena Kulinskaya, David C. Hoaglin, and Ilyas Bakbergenuly}
\date{\today}

\maketitle
\abstract{
This report continues  our investigation of  effects a simulation design may have on the conclusions on  performance of statistical  methods. In the context of meta-analysis of log-odds-ratios,  we consider five generation mechanisms for control probabilities and log-odds-ratios. Our first report (\cite{arXiv_LOR_simulation_equal_sample_sizes}) considered constant sample sizes. Here we report on the results for normally and uniformly distributed sample sizes.
  }

\section{Introduction}
Our interest lies in  effects that simulation design choices may have on conclusions on the comparative merits of various methods, taking as an example meta-analysis of odds ratios.  The basic data from $K$ studies involve $2 K$ binomial variables, $X_{ij}\sim \Bin(n_{ij}, p_{ij})$ for $i = 1,\ldots, K$  and $j = C$ or $T$ (for the Control or Treatment arm); those data underlie the odds-ratios for the meta-analysis.

A design specifies the number of studies, $K$; the sample sizes, $n_{ij}$; the nuisance parameters (control-arm probabilities, $p_{iC}$, or, equivalently, their logits, $\alpha_i$); the overall log-odds-ratio, $\theta$; and the between-study variance, $\tau^2$. For each situation the simulation uses $M$ replications, where $M$ is typically large, say 10,000.

For simplicity, we consider equal arm-level sample sizes, $n_{iC} = n_{iT} = n_i $. The control probabilities $p_{iC}$ or their logits $\alpha_i$ can be constant or generated from some distribution. Normal and uniform distributions are the typical choices. As in our previous report \cite{arXiv_LOR_simulation_equal_sample_sizes}, we consider five possible generation mechanisms for control-arm probabilities and log-odds-ratios under the random-effects model of meta-analysis.

We consider two fixed-intercept random-effects models (FIM1 and FIM2) and two random-intercept random-effects models (RIM1 and RIM2), as in \cite{bakbergenuly2018GLMM}. These models are  equivalent to Models 2 and 4 (for FIM)  and Models 3 and 5 (for RIM), respectively, of \cite{jackson2018comparison}.
Briefly, the FIMs include fixed control-arm effects (log-odds of the control-arm probabilities), and the RIMs replace these fixed effects with random effects. We also consider a model with uniformly distributed control-arm probabilities (URIM1).

Studies also vary in how they specify the sample sizes $n_i$.  In our previous report (\cite{arXiv_LOR_simulation_equal_sample_sizes}) we set $n_1 = \cdots = n_K$ in all $M$ replications. Here we investigate the use of normal and uniform distributions to generate a new set of $n_i$ in each replication.

\section{Generation of sample sizes} \label{sec:GenSampSizes}

Several authors \cite{Cheng2016, bakbergenuly2018GLMM} use constant study-level sample sizes, either equal or unequal, in all replications. More often, however, authors generate sample sizes from a uniform or normal distribution.  \cite{jackson2018comparison}  use (mostly with $n_{iC} = n_{iT}$) sample sizes from discrete $U(50,500)$. \cite{Langan_2018_RSM_1316} use either constant and equal sample sizes within and across studies, or sample sizes from $U(40,\;400)$ and $U(2000,\;4000)$; \cite{sidik2007} use $U(20,\;200)$; and \cite{AboZaid2013} use $U(30,\;100)$ and $U(30,\;1000)$.
\cite{viechtbauer2007confidence}
generates study-level sample sizes ($n_i = n_{iC} = n_{iT}$)  from $N(n,\;n/4)$ ($n/4$ is the variance)  with $n = 10, 20, 40, 80, 160$.
 In an extensive simulation study for sparse data, \cite{kuss2015statistical} uses FIM1 and the corresponding model with $\tau^2 = 0$, along with a large number of fitting methods;  
he generates both the number of studies $K$ and their sample sizes $n$ from log-normal distributions: LN(0.65,\;1.2) and LN(3.05,\;0.97) for $K$ and LN(4.615,\;1.1) for sample sizes.

In general, if mutually independent random variables $Y_i$ have a common distribution $F(\cdot)$, and $N \sim G_n(\cdot)$ is independent of the $Y_i$, the sum $Y_1 + \cdots + Y_N$ has a {\textit compound} distribution \cite{compound}. A binomial distribution with a random number of trials is a compound Bernoulli distribution. The first two moments of such a distribution are $\e(X) = p \e(N)$ and $\var(X) = p(1-p) \e(N) + p^2 \var(N)$. This variance is larger than the variance of the $\Bin(\e(N), \;p)$ distribution. Therefore, random generation of sample sizes produces an overdispersed Binomial (compound Bernoulli) distribution for the control arm, and may also inflate, though in a more complicated way,  the variance in the treatment arm.

In particular, when $N \sim N(\e(N),\sigma^2_n)$, the compound Bernoulli distribution has variance $\var(X) =  p(1 - p) \e(N) + p^2 \sigma^2_n$.
And when $N \sim U(n_l, \;n_u)$,  $\var(X) =  p (1 - p) \e(N) + p^2 (n_u - n_l)^2 / 12$.

\section{Variances  of estimated log-odds-ratios  for random sample sizes}

The (conditional, given $p_{ij}$ and $n_{ij}$) variance of the estimated log-odds-ratio $\hat{\theta}_i$, derived by the delta method, is
\begin{equation} \label{eq:sigma}
v_{i}^2 = {\var}(\hat{\theta}_{i}) = \frac{1} {n_{iT} {p}_{iT} (1 - {p}_{iT})} + \frac{1}{n_{iC} {p}_{iC} (1 - {p}_{iC})},
\end{equation}
estimated by substituting $\hat{p}_{ij}$ for $p_{ij}$.  (We follow the particular method's procedure for calculating $\hat{p}_{ij}$.)

Under the binomial-normal random-effects model (REM), the true study-level
effects, $\theta_i$, follow a normal distribution: $\theta_i \sim N(\theta,\tau^2)$.

To calculate the variance of $\hat{\theta}_i$ when sample sizes $n_i$ are random, we use the law of total variance:
$$\var(\hat\theta_i) = \e(\var(\hat\theta_i | n_i)) + \var(\e(\hat\theta_i | n_i)).$$
The second term is $\var(\theta)=0$,
and the first term is obtained by substituting $\e(n_{iC}^{-1})$ and $\e(n_{iT}^{-1})$ in an expression for the  variance of $\hat\theta$ under fixed sample sizes.

For a random sample size $N$, using the delta method,
\begin{equation} \label{eq:var_infl}
\e(N^{-1}) = (\e(N))^{-1} (1 + [\cv(N)]^2 ),
\end{equation}
where $\cv$ is the coefficient of variation (i.e., the ratio of the standard deviation of $N$  to its mean). Therefore, to order $1/\e(N)$, random generation of sample sizes inflates the variance of $\hat\theta$ if and only if the coefficient of variation of the distribution of sample sizes is of order $1$. In the simulations of \cite{viechtbauer2007confidence}, where $\var(N)=n/4$,  $\cv(N) = O(1/\sqrt{n})$, so the variance is not inflated. In contrast, generating sample sizes from $N(n, n^2/4)$ would result in $\cv=1/2$ and would inflate variance. (Use of such a combination of mean and variance, however, is unlikely to produce realistic sets of sample sizes, and the probability of generating a negative sample size exceeds 2\%.)

The variance of a uniform distribution on an interval of width $\Delta$ centered at $n_0$ is $\Delta^2/12$, and its CV is $\Delta/(\sqrt{12}n_0)$.  Therefore, $\cv(N)$ is of order 1 whenever the width of the interval is of the same order as its center.  Hence, variance is  inflated in simulations by \cite{jackson2018comparison},  \cite{Langan_2018_RSM_1316}, \cite{sidik2007}, and  \cite{AboZaid2013}, who all use wide intervals for $n$.

\section{Design of the simulations for randomly distributed sample sizes}
 \label{sec:design_normally_distributed_sample_sizes}
Our simulations keep the arm-level sample sizes equal and the control-arm probabilities $p_{iC}$ and the log-odds-ratios $\theta_i$ independent. Table~\ref{tab:components} shows the components of the simulations for normally and uniformly distributed sample sizes: parameters, data-generation mechanisms, and estimation targets. Our first report
\cite{arXiv_LOR_simulation_equal_sample_sizes} provides more details. We included the DerSimonian-Laird (DL), restricted maximum-likelihood (REML), Mandel-Paule (MP), and Ku\-lin\-skaya-Dollinger (KD) estimators { of $\tau^2$  with corresponding inverse-variance-weighted estimators of $\theta$ and confidence intervals with critical values from the normal distribution}. \cite{Bakbergenuly2020} studied those inverse-variance-weighted estimators in detail.
We also included the SSW point estimator of $\theta$, whose weights depend only on the studies' arm-level sample sizes, and a corresponding confidence interval, which uses $\hat{\theta}_{SSW}$ as the midpoint, $\hat{\tau}^{2}_{KD}$ in the estimate of its variance, and critical values from the $t$ distribution on $K - 1$ degrees of freedom.  Among the estimators, FIM2 and RIM2 denote the estimators in the corresponding GLMMs.

We generated the arm-level sample sizes, $n_i$, from a normal or a  uniform distribution centered at 40, 100, 250, and 1000.

In generating sample sizes from a  normal distribution, we want negative sample sizes to have reasonably small probability. For our choice of $\sigma^2_n=1.21n^2$ this probability is $0.0008$. Unfortunately,  we were still getting a small number of values below zero out of thousands of  simulated values, so we additionally truncate the $n$ values generated from a normal distribution at 10. Truncation happens with probability $0.009$.

To make uniform distributions of sample sizes  comparable to the normal distributions, we centered them at the same value, $n$, and equated their variances. If a normal distribution has variance $\sigma^2_n$, a uniform distribution with the same variance has interval width $\Delta_n = \sqrt{12 \sigma^2_n}$. We set $\Delta_n = 1.1n$, resulting in $\cv = \Delta_n / (\sqrt{12} n) = 0.318$ and a squared CV  of $0.101$. Therefore, by Equation~(\ref{eq:var_infl}), our simulations with random $n$ inflate variances and covariances by $10\%$ in comparison with simulations with constant $n$. Wider intervals of $n$ would inflate variances more, but in generating sample sizes from a corresponding normal distribution, we wanted negative sample sizes to have reasonably small probability. For our choice of $\Delta_n$ this probability is $0.0008$.


\begin{table}[ht]
	\caption{ \label{tab:components} Components of the simulations for log-odds-ratio}
	\begin{tabular}
		{|l|l|}
		\hline
		Parameter & Values \\
		\hline
		$K$              & 5, 10, 30 \\
		$n$              & 40, 100, 250, 1000  \\
		$\theta$       & 0, 0.5, 1, 1.5, 2 \\
		$\tau^{2}$    & 0(0.1)1 \\
		$p_C$         & .1, .4 \\
		$\sigma^2$  & 0.1, 0.4 \\
        \hline
        Generation of $n$ & \\
        \hline
		Normal($n$, $1.21 n^2 / 12$) & \\
		Uniform($n \pm 0.55n$) & \\
		\hline
		Generation of $p_{iC}$ and $p_{iT}$  & \\
		\hline
		FIM1   & Fixed intercept models: $p_{iC}\equiv p_C$ \\
		FIM2   & \\
		RIM1   & Random intercept models: $\logit(p_{iC})\sim N(\alpha, \sigma^2)$ \\
		RIM2   &  \\
		URIM1 &  $p_{iC} \sim U(p_{C} - \sigma\sqrt{3}p_{C}(1 - p_{C}), p_{C} + \sigma\sqrt{3}p_{C}(1 - p_{C}) )$ \\
		\hline
		Estimation targets & Estimators \\
		\hline
		bias in estimating $\tau^{2}$ & DL, REML, MP, KD, FIM2. RIM2 \\
		bias in estimating $\theta$    & DL, REML, MP, KD, FIM2, RIM2, SSW \\
		coverage of $\theta$             & DL, REML, MP, KD, FIM2, RIM2, \\
		& SSW (with $\hat{\tau}^{2}_{KD}$ and $t_{K - 1}$ critical values) \\
		\hline
	\end{tabular}
\end{table}

\section{Summary of the results}

Our simulations explored two main components of design: the data-generation mechanism and the distribution of study-level sample sizes. Results of our simulations with normally distributed sample sizes are provided in Appendix A, and those with uniformly  distributed sample sizes in Appendix B.

The five data-generation mechanisms (FIM1, FIM2, RIM1, RIM2, and URIM1) often produced different results for at least one of the measures of performance (bias of estimators of $\tau^2$, bias of estimators of $\theta$, and coverage of confidence intervals for $\theta$). In the most frequent pattern FIM2 and RIM2 yield similar results, and FIM1, RIM1, and URIM1 also yield results that are similar but different from those of FIM2 and RIM2. In some situations URIM1 stands apart.

We also expected the coverage of  $\theta$ to suffer because random sample sizes increase the variance of generated log-odds-ratios.  However, generation of sample sizes from normal and uniform distributions had essentially no impact, as can be seen by comparing the results from this report with those from our report  \cite{arXiv_LOR_simulation_equal_sample_sizes} on the simulations with constant sample sizes.
The explanation may lie in our choice of variance $\sigma^2_n$ (not large enough) for the normal and uniform distributions of the sample sizes, causing an increase of just 10\% in the variance of LORs,  or in the rather low coverage, even under constant sample sizes, resulting from considerable biases of estimators of $\theta$.\\

\bibliographystyle{plainnat}
\bibliography{simul_20Feb20}
\end{document}